\documentclass[%
 preprint,
 amsmath,amssymb,
 aps,
]{revtex4-1}

\usepackage{graphicx}
\usepackage{dcolumn}
\usepackage{bm}
\usepackage{hyperref}

\usepackage{amssymb}
\usepackage{amsthm}
\usepackage{comment}

\newtheorem{lemma}{\sc \bf Lemma}[section]
\newtheorem{propos}{\sc \bf Proposition}[section]
\newtheorem{theor}{\sc \bf Theorem}[section]
\newtheorem{corr}{\sc \bf Corollary}[section]
\newtheorem{remark}{\sc \bf Remark}[section]
\newtheorem{definition}{\sc \bf Definition}[section]
\newtheorem{example}{\sc \bf Example}[section]

\newenvironment{cor}{\begin{corr} \hspace{1mm}}{\end{corr}}

\newenvironment{myenumerate}{
\begin{enumerate}
 \setlength{\itemsep}{1pt}
 \setlength{\parskip}{0pt}
 \setlength{\parsep}{0pt}}{\end{enumerate}
}

\begin{document}


\title{Low Algorithmic Complexity Entropy-deceiving Graphs}

\author{Hector Zenil}
\email{hector.zenil@algorithmicnaturelab.org}
\affiliation{Information Dynamics Lab, Unit of Computational Medicine, Department of Medicine Solna, Center for Molecular Medicine, SciLifeLab, Karolinska Institute, Stockholm, Sweden}
\affiliation{Department of Computer Science, University of Oxford, U.K.}
\affiliation{Algorithmic Nature Group, LABoRES, Paris, France}
\homepage{http://algorithmicnature.org/}

\author{Narsis A. Kiani}%
\thanks{First two authors contributed equally}
\affiliation{Information Dynamics Lab, Unit of Computational Medicine, Department of Medicine Solna, Center for Molecular Medicine, SciLifeLab, Karolinska Institute, Stockholm, Sweden}
\affiliation{Algorithmic Nature Group, LABoRES, Paris, France}

\author{Jesper Tegn\'er}%
\affiliation{Biological and Environmental Sciences and Engineering Division, Computer, Electrical and Mathematical Sciences and Engineering Division, King Abdullah University of Science and Technology (KAUST), Thuwal, Kingdom of Saudi Arabia}
\affiliation{Unit of Computational Medicine, Department of Medicine Solna, Center for Molecular Medicine, SciLifeLab, Karolinska Institute, Stockholm, Sweden}

\date{\today}
\begin{abstract} In estimating the complexity of objects, in particular of graphs, 
it is common practice to rely on graph- and information-theoretic measures. Here, using integer sequences with properties such as Borel normality, we explain how these measures are not independent of the way in which an object, such as a graph, can be described or observed. From observations that can reconstruct the same graph and are therefore essentially translations of the same description, we will see that when applying a computable measure such as Shannon Entropy, not only is it necessary to pre-select a feature of interest where there is one, and to make an arbitrary selection where there is not, but also that more general properties, such as the causal likelihood of a graph as a measure (opposed to randomness), can be largely misrepresented by computable measures such as Entropy and Entropy rate. We introduce recursive and non-recursive (uncomputable) graphs and graph constructions based on these integer sequences, whose different lossless descriptions have disparate Entropy values, thereby enabling the study and exploration of a measure's range of applications and demonstrating the weaknesses of computable measures of complexity.
\end{abstract}

\maketitle


\section{The use of Shannon Entropy in network profiling}

One of the major challenges in modern physics is to provide proper and suitable representations of network systems for use in fields ranging from physics~\cite{boccaletti} to chemistry~\cite{chen2014entropy}. A common problem is the description of order parameters with which to characterize the `\textit{complexity of a network}'. Graph complexity has traditionally been characterized using graph-theoretic measures such as degree distribution, clustering coefficient, edge density, and community or modular structure. 

More recently, networks have also been characterized using classical information theory. One problem in this area is the interdependence of many graph-theoretic properties, which makes measures more sophisticated than single-property measurements~\cite{orsini} difficult to come by. The standard way to address this is to generate graphs that have a certain specific property while being random in all other aspects, in order to check whether or not the property in question is typical among an ensemble of graphs with otherwise seemingly different properties. 

Approaches using measures based upon Shannon Entropy's claim to quantify the information content of a network~\cite{bianconi2007entropy} as an indication of its `typicality' are based on an assumption of associated ensembles provided by the Entropy evaluation: the more random the more typical. The claim is that one can construct a ``null model'' that captures some aspects of a network (e.g. graphs that have the same degree distribution) and see how different the network is to the null model as regards particular features, such as clustering coefficient, graph distance, or other features of interest. The procedure aims at producing an intuition of an ensemble of graphs that are assumed to have been sampled uniformly at random from the set of all graphs with the same property to determine if such a property occurs with high or low probability. If the graph is not significantly different, statistically, from the null model, then the graph is claimed to be as ``simple'' as the null model; otherwise, the measure is claimed to be a lower bound on the ``complexity'' of the graph as an indication of its random versus causal nature. 

Here we highlight some serious limitations of these approaches that are often neglected, and provide pointers to approaches that are complementary to Shannon Entropy, in order to partially circumvent some of the aforesaid limitations by combining it with a measure of local algorithmic complexity that better captures the recursive and thus causal properties of an object--in particular a network--beyond statistical properties. 

One of the most popular applications of Entropy is to graph degree distribution, as first suggested and introduced by~\cite{korner1988random}. Similar approaches have been adopted in areas such as chemical graph theory and computational  systems biology~\cite{dehmer2008entropy} as functions of layered graph degree distribution under certain layered coarse-graining operations (sphere covers), leading to the hierarchical application of Entropy, a version of graph traversal Entropy rate. In chemistry, for example, Shannon Entropy over a function of degree sequence has been used as a profiling tool to characterize--so it is claimed--molecular complexity. 

While the application of Entropy to graph degree distributions has been relatively more common, the same Entropy has also been applied to other graph features, such as functions of their adjacency matrices~\cite{estrada2014walk}, and to distance and Laplacian matrices~\cite{Dehmer3}. 

Even more recently, Shannon Entropy on adjacency matrices was used to attempt the discovery of CRISPR regions in an interesting transformation of DNA sequences into graphs~\cite{sengupta2016application}. A survey contrasting adjacency matrix based (walk) entropies and other entropies (e.g. on degree sequence) is offered in~\cite{estrada2014walk}. It finds that adjacency based ones are more robust vis-a-vis graph size and are correlated to graph algebraic properties, as these are also based on the adjacency matrix (e.g. graph spectrum). 

Finally, hybrid measures have been used, such as the graph heterogeneity index~\cite{Estrada2} as a function of degree sequence, and the Laplacian matrix, where some of the limitations of quantifying only the diversity of the degree distribution, i.e. its Entropy (or of any graph measure as a  function of the Entropy of the degree distribution), have been identified. 

It is thus of the greatest interest to researchers in physics, chemistry and biology to understand the reach, limits and interplay of measures of entropy, in particular as applied to networks. Likewise to understand how unserviceable for extracting causal content--as opposed to randomness--the use of entropy as a measure of randomness, complexity or information content can be. The use of entropy has, however, been extended, because its numerical calculation is computationally very cheap as compared to richer, but more difficult to approximate universal measures of complexity which are better qualified to capture more general properties of graphs. Some of these properties to be captured are related to the nature of the graph-generating mechanisms, which were what most of the previously utilized measures were supposed to quantify in the first place, in one way or another, from the introduction of the first random graph model by Erd{\"o}s and R\'enyi~\cite{erdos1960evolution} to the most popular models such as `scale-freeness'~\cite{barabasi}, and more recent ones such as network randomness typicality~\cite{bianconi2007entropy}.

\section{Notation and basic definitions}

\begin{definition}
A graph is an ordered pair $G = (V,E)$ comprising a set $V$ of nodes or vertices and a set $E$ of edges or links, which are 2-element subsets of $V$.
\end{definition}

\begin{definition}
A graph $G$ is \textit{labelled} when the vertices are distinguished by labels $u_1, u_2, \ldots u_n$, with $n=|V(G)|$ the cardinality of the set $V(G)$.
\end{definition}

\begin{definition}
Graphs $G$ and $H$ are \textit{isomorphic} if there is a bijection between the vertex sets of $G$ and $H$, $\lambda : V(G) \rightarrow V(H)$ such that any two vertices $u$ and $v \in V(G)$ are adjacent in $G$ if and only if $\lambda(u)$ and $\lambda(v)$ are adjacent in $H$.
\end{definition}

\begin{definition}
The \textit{degree} of a node $v$, denoted by $d(v)$, is the number of (both incoming and outgoing) links to other nodes, and $d$ is the unordered list of all $v \in V(G)$.
\end{definition}

\begin{definition}
An E-R graph $G(n, p)$ is a graph of size $n$ constructed by connecting nodes randomly with probability $p$ independent of every other edge.
\end{definition}

Usually E-R graphs are assumed to be non-recursive (i.e. truly random), but E-R graphs can be constructed recursively using pseudo-random generating algorithms.

\subsection{Graph Entropy}

One of the main objectives behind the application of Shannon Entropy is the characterization of the randomness or `information content' of an object such as a graph. Here we introduce graphs with interesting deceptive properties, particularly disparate Entropy (rate) values for the same object when looked at from different perspectives, revealing the inadequacy of classical information-theoretic approaches to graph complexity.

Central to information theory is the concept of Shannon's information Entropy, which quantifies the average number of bits needed to store or communicate the statistical description of an object.

For an ensemble $X(R, p(x_i))$, where $R$ is the set of possible outcomes (the random variable), $n=|R|$ and $p(x_i)$ is the probability of an outcome in $R$. The Shannon Entropy of $X$ is then given by

\begin{definition}
\begin{equation}
H(X)=-\sum_{i=1}^n p(x_i) \log_2 p(x_i)
\end{equation}
\end{definition}

Which implies that to calculate $H(X)$ one has to know or assume the mass distribution probability of ensemble $X$. One caveat regarding Shannon's Entropy is that one is forced to make an arbitrary choice regarding granularity. Take for example the bit string 01010101010101. The Shannon Entropy of the string at the level of single bits is maximal, as there are the same number of 1s and 0s, but the string is clearly regular when 2-bit (non-overlapping) blocks are taken as basic units, in which instance the string has minimal complexity because it contains only 1 symbol (01) from among 4 possible ones (00,01,10,11). A generalization consists in taking into consideration all possible ``granularities'' or the Entropy rate:

\begin{definition}
Let $Pr(s_i, s_{i+1}, \ldots, s_{i+L}) = Pr(s)$ with $|s|=L$ denote the joint probability over blocks of $L$ consecutive symbols. Let the \textit{Shannon Entropy rate}~\cite{shannon} (also known as granular Entropy, $n$-gram Entropy) of a block of $L$ consecutive symbols--denoted by $H(L)$--be:
\begin{equation}
H_L(s) = -\sum_{s1\in A} \ldots \sum_{s_L\in A} Pr(s_1, \ldots, s_L) \log_2 Pr(s_1, \ldots, s_L)
\end{equation}
\end{definition}

Thus to determine the Entropy rate of the sequence, we estimate the limit when $L \rightarrow \infty$. It is not hard to see, however, that $H_L(s)$ will diverge as $L$ tends to infinity if the number of symbols increases, but if applied to a binary string $H_L(s)$, it will reach a minimum for the granularity in which a statistical regularity is revealed. 

The Shannon Entropy~\cite{shannon} of an object $s$ is simply $H_L(s)$ for fixed block size $L=i$, so we can drop the subscript. 

We can define the Shannon Entropy of a graph $G$, with respect to $i$, by:

\begin{definition}
\begin{equation}
H(G,P)= -\sum_i^{|G|} P(G_i) \log_2 P(G_i)
\end{equation}
\end{definition}

\noindent where $P$ is a probability distribution of $G_i$, $i$ is a feature of interest of $G$, e.g. edge density, degree sequence, number of over-represented subgraphs/graphlets (graph motifs), and so on. When $P$ is the uniform distribution (every graph of the same size is equally likely), it is usually omitted as a parameter of $H$.

The most common applications of Entropy to graphs are to degree sequence distribution and edge density (adjacency matrix), which are labelled graph invariants. In molecular biology, for example, a common application of Entropy is to count the number of `branchings'~\cite{mowshowitz} per node by, e.g., randomly traversing a graph starting from a random point. The more extensive the branching, the greater the uncertainty of a graph's path being traversed in a unique fashion, and the higher the Entropy. Thorough surveys of graph Entropy are available in~\cite{mowshowitz,simonyi,mowshowitz2}, so we will avoid providing yet another one. In most, if not all of these applications of Entropy, very little attention is paid to the fact that Entropy can lead to completely disparate results depending on the ways in which the same objects of study are described, that is, to the fact that Entropy is not a graph invariant--either for labelled or unlabelled graphs--vis-\'a-vis object description, a major drawback for a complexity measure~\cite{zenildata,zenilbdm} of typicality, randomness, and causality. In the survey~\cite{mowshowitz}, it is suggested that there is no `right' definition of Entropy. Here we formally confirm this to be the case in a fundamental sense. 

Indeed, Entropy requires a pre-selection of a graph invariant, but it is itself not a graph-invariant. This is because ignorance of the probability distribution makes Entropy necessarily dependent on graph invariant description, there being no such thing as an \textit{Invariance theorem}~\cite{solomonoff,kolmogorov,chaitin} in Shannon Entropy to provide a convergence of values independent of description language as there is in algorithmic information theory for algorithmic (Kolmogorov-Chaitin) complexity. 

\begin{definition}
The algorithmic complexity of an object $G$ is the length of its shortest computational description (computer program) in a reference language (of which it is independent), such that the shortest generating computer program fully reconstructs $G$~\cite{solomonoff,kolmogorov,chaitin, levin}.
\end{definition}

\section{Construction of Entropy-deceiving graphs}

If we can show that we can artificially fool Entropy we will show how Entropy may fail to characterize natural or socially occurring networks. Especially because, as we will demonstrate, different values of Shannon Entropy can be retrieved for the same graph as functions of different features of interest of  said graph, thereby showing that there is no such thing as the `Shannon Entropy of a graph' but rather `the Shannon Entropy of an \textit{identified property} of a graph', which can easily be replaced by a function that simply quantifies such a property directly.

\subsection{Entropy of pseudo-random graphs} By using integer sequences, in particular Borel-normal irrational numbers,

 one can construct pseudo-random graphs, which can in turn be used to construct networks. 

\begin{definition}
A real number $x$ is said to be normal if all $n$-tuplets of $x$'s digital expansion are equally likely, thereby of natural maximal $n$-order Entropy rate by definition of Borel normality.
\end{definition}

For example, the mathematical constant $\pi$ is believed to be an absolute Borel normal number (Borel normal in every base), and so one can take the digits of $\pi$ in any base and take $n \times n$ digits as the entries for a graph adjacency matrix of size $n \times n$ by taking $n$ consecutive segments of $n$ digits $\pi$. The resulting graph will have $n$ nodes and an edge density 0.5 because the occurrence of 1 or 0 in $\pi$ in binary has the probability 0.5 (the same as $\pi$ in decimals after transformation of digits to 0 if digit $i<5$ and 1 otherwise, or $i<b/2$ and 1 otherwise in general for any base $b$), thus complying with the definition of an Erd{\"o}s-R\'enyi (E-R) graph (albeit of high density).

\begin{figure}[ht!]
\centering

\scalebox{.27}{\includegraphics{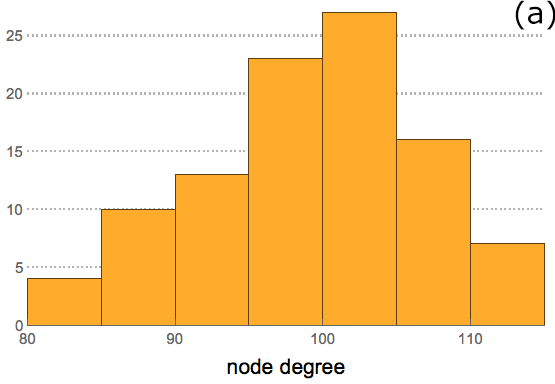}}\hspace{1cm}\scalebox{.3}{\includegraphics{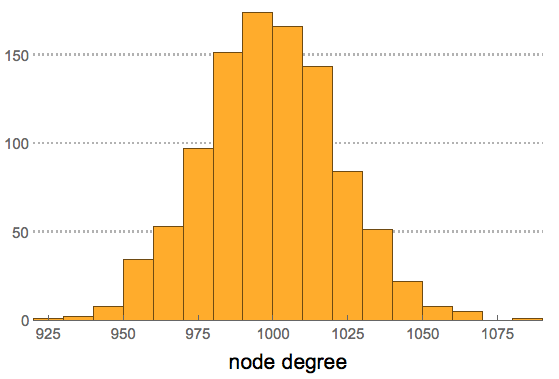}}\\

\bigskip

\medskip

\scalebox{.27}{\includegraphics{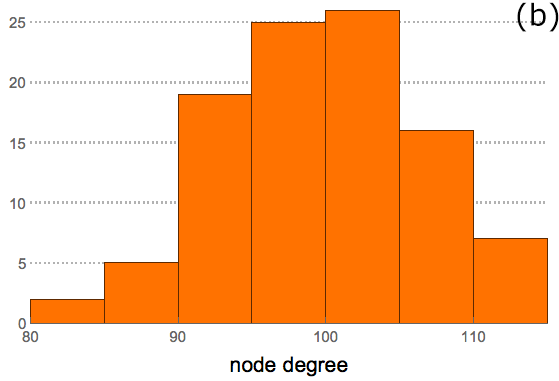}}\hspace{0.8cm}\scalebox{.3}{\includegraphics{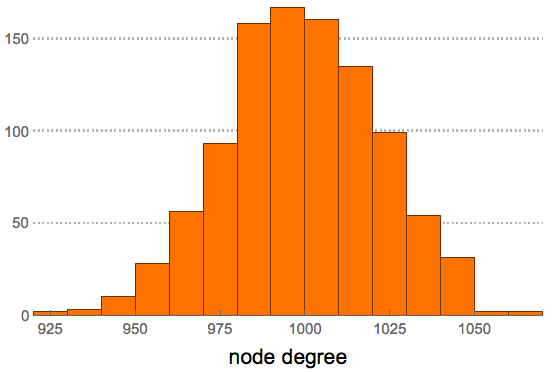}}\\

\bigskip

\medskip

\scalebox{.34}{\includegraphics{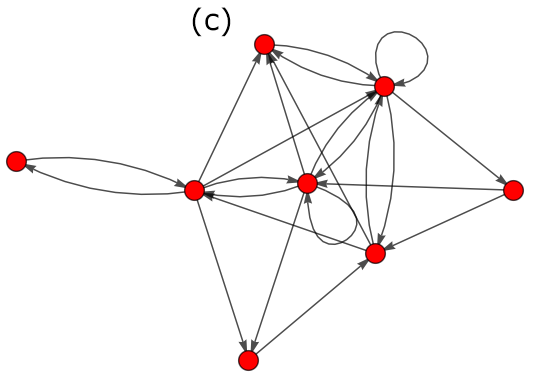}}
\caption{\label{pigraphs}Histograms of degree distributions of $\pi$ networks using 10\,000 (left) and $10\times10^6$ (right) digits of $\pi$ in base 2 (A) and in base 10 (B) undirected and with no self-loops. C: A graph based on the 64 calculated bits of a partially computable $\Omega$ Chaitin number~\cite{calude}. It appears to have some structure but any regularity will eventually vanish as it is a Martin-L{\"o}f algorithmic random number~\cite{martinlof}.}
\end{figure}

As theoretically predicted and numerically demonstrated in Fig.~\ref{pigraphs}(A and B), the degree distribution will approximate a normal distribution around $n$. This means that the graph adjacency matrix will have maximal Entropy (if $\pi$ is Borel normal) but low degree-sequence Entropy because all values are around $n$ and they do not span all the possible node degrees (in particular, low degrees). This means that algorithmically constructing a graph can give rise to an object with a different Entropy when the feature of interest of the said graph is changed. 

A graph does not have to be of low algorithmic complexity to yield incompatible observer-dependent Entropy values. One can take the digits of an $\Omega$ Chaitin number (the halting probabilities of optimal Turing machines with prefix-free domains), some of the digits of which are uncomputable. But in Fig.~\ref{pigraphs}(C) we show a graph based on the first 64 digits of an $\Omega$ Chaitin number~\cite{calude}, thus a highest-algorithmic-complexity graph in the long run (it is ultimately uncomputable). Since randomness implies normality~\cite{martinlof}, the adjacency matrix has maximal Entropy, but for the same reasons as obtain in the case of the $\pi$ graphs, it will have low degree-sequence Entropy. For algorithmic complexity, in contrast, as we will see in Theorem~\ref{algo}, all graphs have the same algorithmic complexity regardless of their (lossless) descriptions (e.g. adjacency matrix or degree sequence), as long as the same and only the same graph (up to an isomorphism) can be reconstructed from their descriptions.

\begin{figure}[ht!]
\centering
\scalebox{.30}{\includegraphics{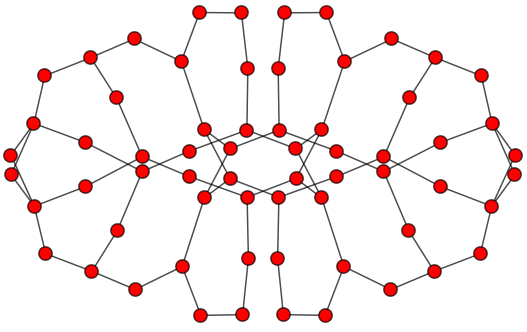}}\hspace{1.5cm}\scalebox{.33}{\includegraphics{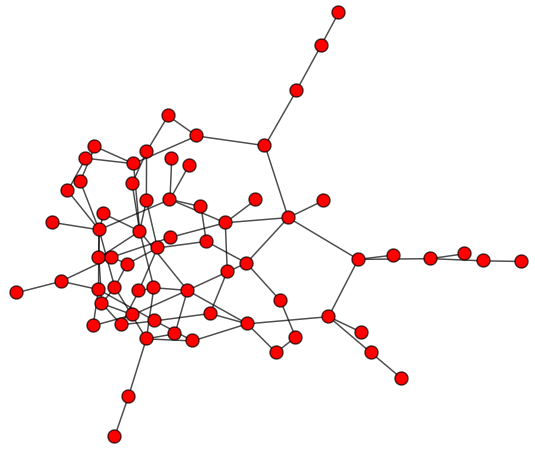}}
\caption{\label{Entropy}A regular \textit{antelope} graph (left) and an Erd\"os-R\'enyi (E-R) graph (right) with the same number of edges and nodes, therefore the same adjacency matrix dimension and exactly the same edge density $0.03979\ldots$ can have very different properties. Specifically, one can be recursively (algorithmically) generated while the other is random looking. One would wish to capture this essential difference.}
\end{figure}

One can also start from completely different graphs. For example, Fig.~\ref{Entropy} shows how Shannon Entropy is applied directly to the adjacency matrix as a function of edge density, with the same Entropy values retrieved despite their very different (dis)organization. 

\begin{figure}[ht!]
\centering
\scalebox{.43}{\includegraphics{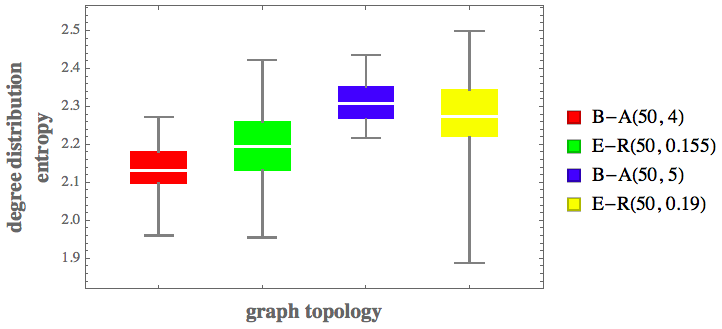}}\\
\caption{\label{distEntropy}Box plot of Entropy values applied to the degree-sequence distribution of 10 scale-free (B-A) and 10 E-R graphs with $n=50$ nodes and the same parameters. Results may mislead as to the generative quality of each group of graphs, suggesting that B-A are as or more random than E-R graphs, despite their recursive (causal/algorithmic and deterministic) nature, whereas in fact this should make B-A networks more random than E-R graphs. Here the E-R graphs have exactly the same edge density as the B-A graphs for 4 and 5 preferential attached edges per node. This plot illustrates how, for all purposes, Entropy can be easily fooled and cannot tell apart higher causal content from apparent randomness. One can always update the ensemble distribution to accommodate special cases but only after gaining knowledge by other methods.}
\end{figure}

The Entropy rate will be low for the regular antelope graph, and higher, but still far removed from randomness for the E-R, because by definition the degree-sequence variation of an E-R graph is small. However, in scale-free graphs degree distribution is artificially scaled, spanning a large number of different degrees as a function of number of connected edges per added node, and resulting in an over-estimation of their degree-sequence Entropy, as can be numerically verified in Fig.~\ref{distEntropy}. Degree-sequence Entropy points in the opposite direction to the entropic estimation of the same graphs arrived at by looking at their adjacency matrices, when in reality, scale-free networks produced by, e.g., Barabasi-Albert's preferential attachment algorithm~\cite{barabasi}, are recursive (algorithmic and deterministic, even if probabilities are involved), as opposed to the E-R construction built (pseudo-)randomly. The Entropy of the degree-sequence of scale-free graphs would suggest that they are almost as, or even more random than E-R graphs for exactly the same edge densities. To circumvent this, ad-hoc measures of modularity have been introduced~\cite{sole}, to precisely capture how removed a graph is from `scale-freeness' by comparing any graph to a scale-free randomized version of itself, and thereby compelling consideration of a pre-selected feature of interest (`scale-freeness'). 

Furthermore, an E-R graph can be recursively (algorithmically) generated or not, and so its Shannon Entropy has no connection to the causal, algorithmic information content of the graph, and can only provide clues for low Entropy graphs that can be characterized by other graph-theoretic properties, without need of an entropic characterization.

\subsection{A low complexity and high Entropy graph}

\begin{figure}[ht!]
\centering
\scalebox{.26}{\includegraphics{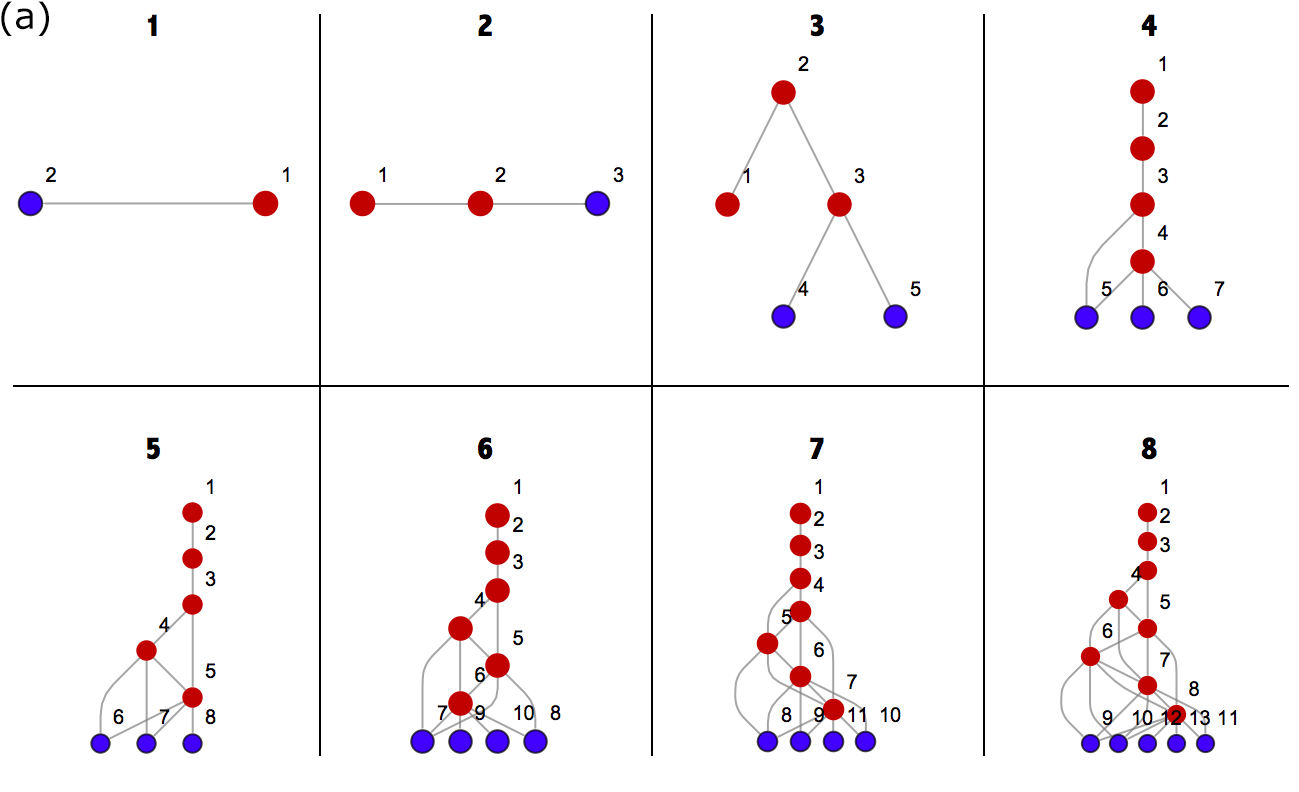}}\\

\medskip
\bigskip
\scalebox{.26}{\includegraphics{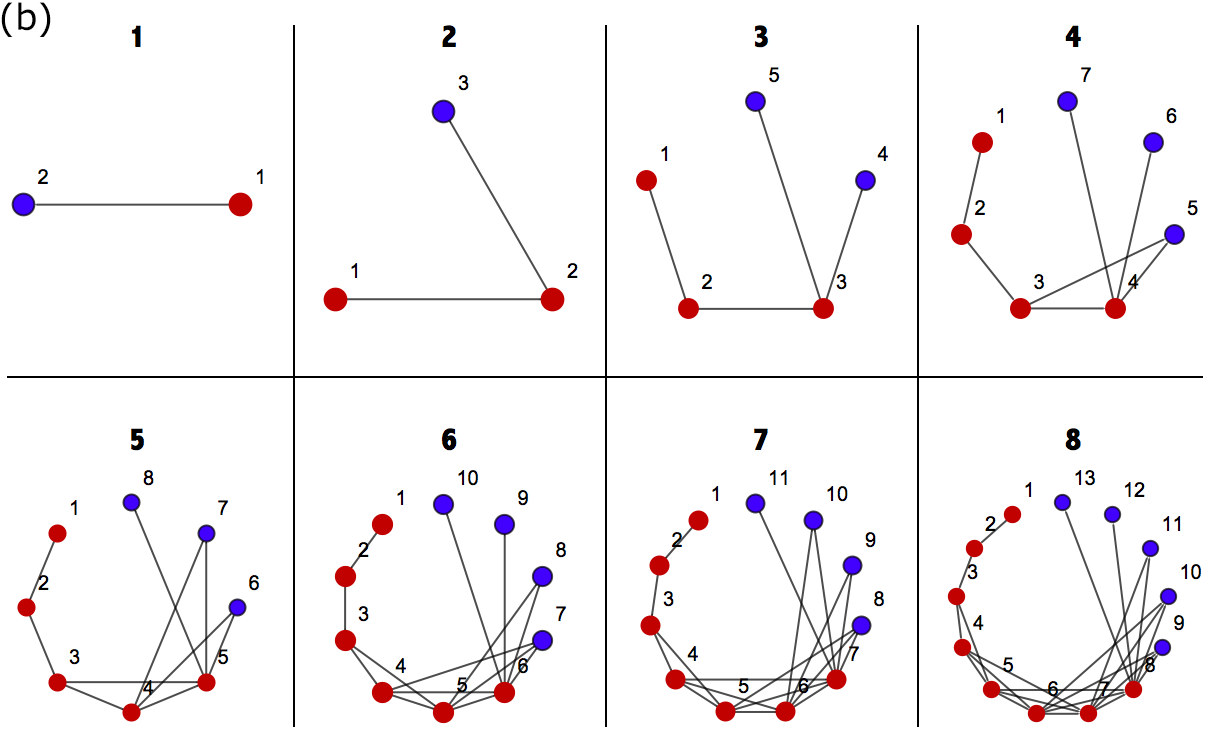}}
\caption{\label{ZKgraph}. Tree-like (a) and radial representation (b) of the same ZK graph with maximal Entropy degree sequence by construction, starting from iteration 2 and proceeding to 8, adding a node at a time.}
\end{figure}

We introduce a method to build a family of recursive graphs with maximal Entropy but low algorithmic complexity, hence graphs that appear statistically random but are, however, of low algorithmic randomness and thus causally (recursively) generated. Moreover, these graphs may have maximal Entropy for some lossless descriptions but minimal Entropy for other lossless descriptions of exactly the same objects, with both descriptions characterizing the same object and only that object, thereby demonstrating how Entropy fails at unequivocally and unambiguously characterizing a graph independent of a particular feature of interest. We denote by `ZK' the graph (unequivocally) constructed as follows:

\begin{myenumerate}
\item Let $1\rightarrow 2$ be a starting graph $G$ connecting a node with label 1 to a node with label 2. If a node with label $n$ has degree $n$, we call it a \textit{core node}; otherwise, we call it a \textit{supportive node}.
\item Iteratively add a node $n+1$ to $G$ such that the number of core nodes in $G$ is maximized. The resulting graph is typified by the one in Fig.~\ref{ZKgraph}.
\end{myenumerate}

\subsection{Properties of the ZK graph}

The degree sequence $d$ of the labelled nodes $d=1, 2, \ldots, n$ is the Champernowne constant~\cite{champernowne} $C$ in base 10, a transcendental real whose decimal expansion is Borel normal~\cite{borel}, constructed by concatenating representations of successive integers 

$$C_{10} = 0.1234567891011121314151617181920\ldots$$ 

whose digits are the labelled node degrees of $G$ for $n=20$ iterations (sequence A033307 in the OEIS).

The sequence of edges is a recurrence relation built upon previous iteration values between core and supportive nodes, defined by:

$$[1/r]+[2/r]+ \ldots +[n/r]$$

\noindent where $r = (1+sqrt(5))/2$ is the golden ratio and $[ \textit{ }]=$ the floor function (sequence A183136 in the OEIS) whose values are 1, 2, 4, 7, 10, 14, 18, 23, 29, 35, 42, 50, 58, 67, 76, 86, 97, 108, 120, 132, 145, $\ldots$ 

\begin{figure}[ht!]
\centering
\scalebox{.4}{\includegraphics{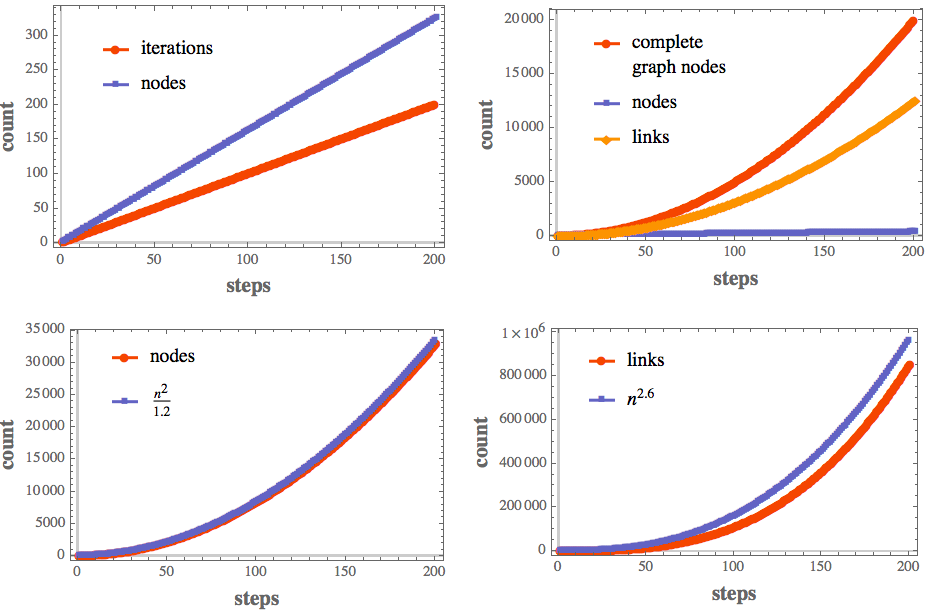}}
\caption{\label{basic}Basic node and link growth properties and corresponding fitted (polynomial) lines. The relation between node and link growth determines the edge density, which at the limit is 0.}
\end{figure}

\begin{figure}[ht!]
\centering
\scalebox{.35}{\includegraphics{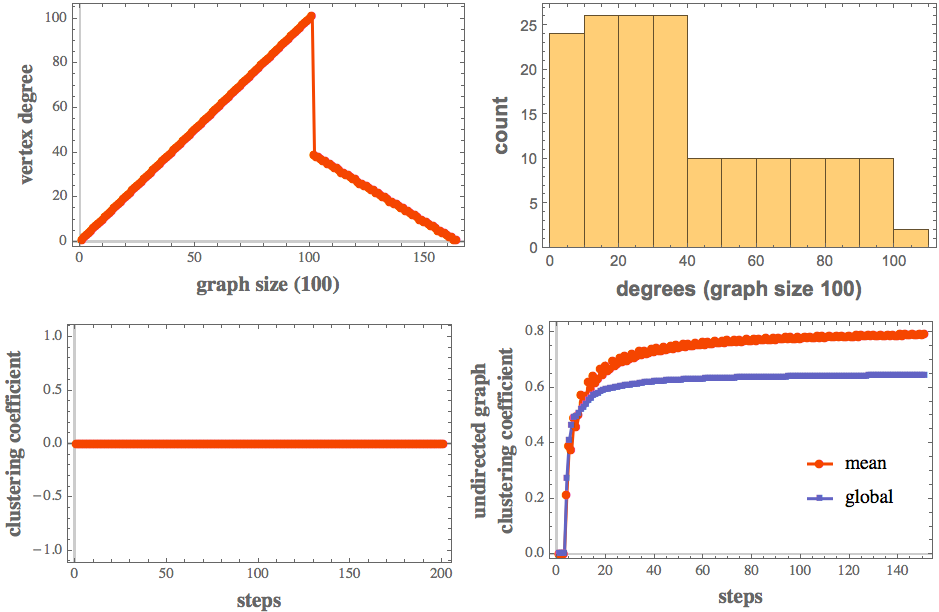}}
\scalebox{.36}{\includegraphics{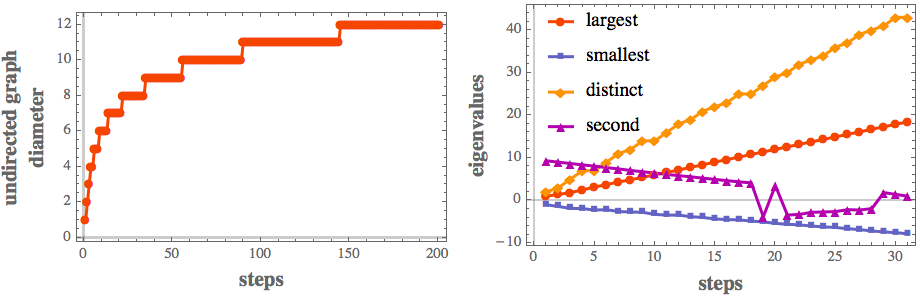}}
\caption{\label{graphtheoretic}Graph theoretic and dynamic properties of the recursive `ZK' graph. Despite the trivial construction of the recursive network, it displays all sorts of interesting convergent and divergent non-trivial graph-theoretic, dynamic and complexity properties. For example, the clustering coefficient of the undirected graph asymptotically converges to 0.65 and some properties grow or decrease linearly while others do so polynomially. Entropy of different graph descriptions (even for fully accurate descriptions, and not because of a lack of information from the observer point of view) diverge and become trivially dependent on other simple functions (e.g. edge density or degree sequence normality). 
In contrast, methods based on algorithmic probability (c.f.~\ref{conclusion}) assign lower complexity to the graph than both Entropy and lossless compression algorithms (e.g. Compress, depicted here) that are based on Entropy rate (word repetition). While useful for quantifying specific features of the graph that may appear interesting, no graph-theoretic or entropic measure can account for the low (algorithmic) randomness and therefore (high) causal content of the network.}
\end{figure}

\begin{figure}[ht!]
\centering
\scalebox{.35}{\includegraphics{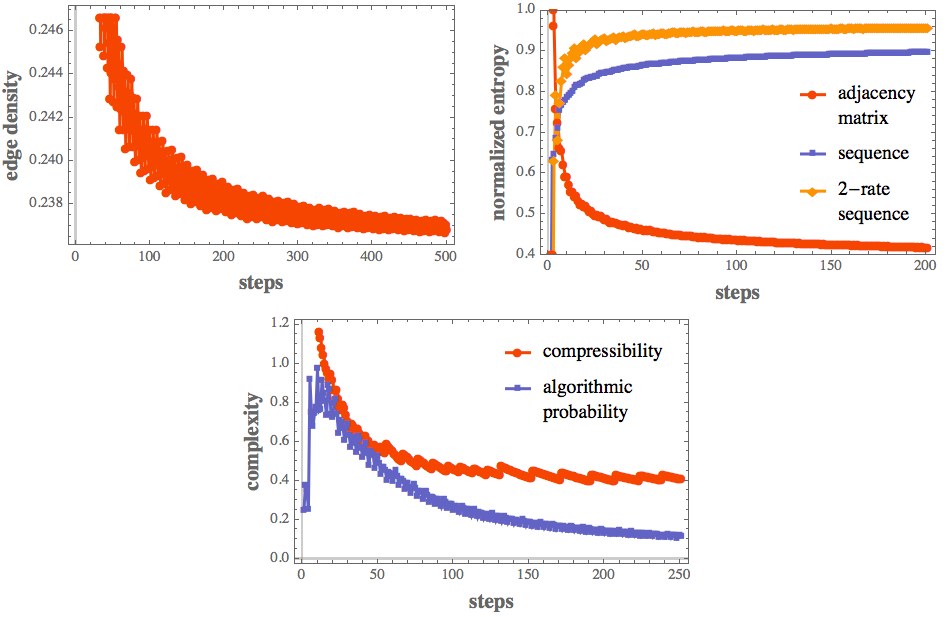}}
\caption{\label{complexity}ZK randomness and information content according to lossless compression Entropy and a technique, other than compression, that uses the concept of algorithmic probability to approximate algorithmic complexity~\cite{kolmo2d,zenilgraph,zenilmethodsbiology}. This means that randomness characterizations by algorithmic complexity are robust, as they are independent of object description, and are therefore, in an essential way, parameter-free, meaning that there is no need for pre-selection or arbitrary selection of features of interest for proper graph profiling.}
\end{figure}
	
\begin{definition}\label{d1}
$ZK^n$ is a graph with at least one node with degree $x$ where  $\forall x \in {1 \cdots n} $.  
\end{definition}

$ZK_m^n$ has been used where we want to emphasize the number of generation- or time-steps in the process of constructing $ZK^n$. The symbol $\Delta(ZK)$ denotes the maximum degree of the graph. 
Nodes in the ZK graph belong to 2 types: core and supportive nodes.
\begin{definition}\label{d2}
Node $x$ is a core node  iff $\exists \: m\in\{1 \cdots n-3\}$ such that $x\:\in \Delta(ZK^n_m)$. Otherwise it is a supportive node.
\end{definition}

\begin{theor}\label{T1}
To convert $ZK^{r-1}$ to $ZK^r$, we need to add 2 supportive nodes to $ZK^{r-1}$  if $r$ is odd or one supportive node if $r$ is even. 
\begin{proof}
\noindent \textit{By induction:}\\

\noindent \textbf{The basis}: $ZK^3$ has 3 core nodes denoted by $c^3$ and 2 supportive nodes denoted by $s^3$. As described in the construction procedure, to convert $ZK^3$ to $ZK^4$, we choose a supportive node with maximum degree. Here, since we have only $s^3$ nodes, their degree is one. So we need to connect to 3 other supportive nodes. As we have only one left, we need to add 2 supportive nodes. Now, $ZK^4$ has 3 supportive nodes, 2 of them new, $s^4$, and one old, $s^3$. The old one is of degree 2, and we need to convert it to 5; we have 2 other supportive nodes left, so we need a new supportive node $s^4$. Therefore, the assumption is true for $ZK^3$ and $ZK^4$ (the basis).

\noindent \textbf{Inductive step}: Now, if we assume that it is true for $ZK^{n-1}$, then it is true for $ZK^n$.\\

\noindent We consider 2 cases:

\begin{myenumerate}
\item  $n-1$ is odd
\item $n-1$ is even
\end{myenumerate}
 
\noindent \textbf{Case one:}
If $n-1$ is odd then $n-2$ is even, which means we have added one supportive node with degree one, and to convert $ZK^{n-1}$ to $ZK^n$ we need to have a core node with degree $n$. The maximum degree of a supportive node is $n-3$, and we have only one supportive node which is not connected to the core candidate node, which implies that the core candidate node will be $n-2$, and we would need to add 2 extra supportive nodes to our graph.\\

\noindent \textbf{Case two:}
If $n-1$ is even then $n-2$ is odd, and therefore $ZK^{n-1}$ has 2 supportive nodes with degree one (they have only been connected to the last core nodes). So we would need to add only one node to convert the supportive node with maximum degree to a core node with degree $n$.
\end{proof}
\end{theor}

\begin{cor}
\begin{myenumerate}						\item  if $n$ is odd then  $|V(ZK^n)|= 2n-2$
						 \item $n-1$ is even $|V(ZK^n)|= 2n-1$
\end{myenumerate}
 
 \end{cor} 
 
\begin{theor}\label{T4}
$\forall r \in {1\cdots n}$ there is a maximum of 3 nodes with degree $r$ in $ZK^n$.
\begin{proof}
\noindent \textit{By induction:}\\

\noindent \textbf{The basis}: The assumption is true for  $ZK^3$. \\

\noindent \textbf{Inductive step}: If we assume $ZK^{n-1}$ have $\forall r \in {1\cdots n-1}$ there is a maximum of 3 nodes with degree $r \in ZK^{n-1}$ then $\forall r \in {1\cdots n}$, then there is a maximum of 3 nodes with degree $r \in ZK^n$.

The proof is direct using theorem~\ref{T1}. To generate $ZK^n$, we add a maximum of 2 supportive nodes. These nodes have degree one and there is no node with degree one except the first core node (core node with degree 1). Thus we have a maximum of 3 nodes with degree one. The degree of all other supportive nodes will be increased by one, which, based on the hypothesis of induction, has not been repeated more than 3 times.
\end{proof}
\end{theor}

\begin{theor}\label{maxEntropy}
ZK is of maximal degree-sequence Entropy. 
\begin{proof}
The degree sequence of the ZK graph can be divided into 2 parts:

\begin{myenumerate}
\item A dominating degree subsequence associated with the core nodes (always longer than subsequence 2 of supporting nodes) generated by the infinite series:
$$C_{10}=\sum _{n=1}^{\infty }\sum _{k=10^{n-1}}^{10^{n}-1}{\frac {k}{10^{n(k-10^{n-1}+1)+9\sum _{\ell =1}^{n-1}10^{\ell -1}\ell }}}$$
\noindent that produces the Champernowne constant $C_{10}$, which is Borel normal~\cite{borel,champernowne}.
\end{myenumerate}

\begin{myenumerate}
\item A second degree sequence associated with the supportive nodes, whose digits do not repeat more than 3 times, and therefore, by Theorem~\ref{T4}, has a maximal $n$-order Entropy rate for $n>2$ and a high Entropy rate for $n<2$.
\end{myenumerate}

Therefore, the degree sequence of ZK is asymptotically of maximal Entropy rate.
\end{proof}
\end{theor}

\begin{theor}\label{lowK}
The ZK graph is of low algorithmic (Kolmogorov-Chaitin) complexity.
\begin{proof}
By demonstration: The computer generated program of the ZK graph written in the Wolfram Language, is:

\begin{verbatim}
AddEdges[graph_] := 
 EdgeAdd[graph, 
  Rule@@@Distribute[{Max[VertexDegree[graph]] + 1, 
     Table[i, {i, (Max[VertexDegree[graph]] + 
         2), (Max[VertexDegree[graph]] + 
          1) + (Max[VertexDegree[graph]] + 1) - 
        VertexDegree[graph, Max[VertexDegree[graph]] + 1]}]}, List]]
\end{verbatim}

The graph can be constructed recursively for any number of nodes $n$ by nesting the \texttt{AddEdges[]} function as follows:

\begin{verbatim}
Nest[AddEdges, Graph[{1 -> 2}, n]
\end{verbatim}

\noindent starting from the graph defined by $1 \rightarrow 2$ as initial condition.

The length of \texttt{NestList} with \texttt{AddEdges} and the initial condition in bytes is the algorithmic complexity of ZK, which grows by only $\log_10 i$ and is therefore of low algorithmic randomness.
\end{proof}
\end{theor}

We now show that we can fully reconstruct ZK from the degree sequence. As we know that we can also reconstruct ZK from its adjacency matrix (denoted by $Adj(ZK)$), we therefore have it that both are lossless descriptions from which ZK can be fully reconstructed and for which Entropy provides contradictory values depending on the feature of interest.

\begin{theor}\label{T5}
$\forall n \in \mathbb{N}$, all instances of $ZK^n$ are isomorphic.
\begin{proof}
The only degree of freedom in the graph reconstruction is the selection of a \textit{supportive node} to convert to a \textit{core node} when there are several supportive nodes of maximal degree. As has been proven in Theorem~\ref{T1}, the number of nodes which are added to a graph is independent of the supportive nodes selected for conversion to a core node. In any instance of a graph the number of nodes and edges are equal, and it is clear that by mapping the selected node in each step in any instance of a graph to the selected node in the corresponding step in another instance ${ZK^\prime}^n)$ we get $f: V(ZK^n) \Rightarrow V({ZK^\prime}^n)$, such that $f$ is a bijection (both one-one and superimposed one on the other).
\end{proof}
\end{theor}

Finally, we prove that all isomorphic graphs have about the same (e.g. low) algorithmic complexity:

\begin{theor}\label{algo}
Let $G^\prime$ be an \textit{isomorphic graph} of $G$. Then $K(G^\prime)\sim K(G)$ for all $K(G^\prime) \in Aut(G)$, where $Aut(G)$ is the automorphism group of $G$.

\begin{proof}
The idea is that if there is a significantly shorter program $p^\prime$ for generating $G$ compared to a program $p$ generating $Aut(G)$, we can use $p^\prime$ to generate $Aut(G)$ via $G$ and a relatively short program  $c$ that tries, e.g., all permutations, and checks for isomorphism. Let's assume that there exists a program $p^\prime$ such that $||p^\prime|- |p||>c$, i.e. the difference is not bounded by any constant, and that $K(G)=|p^\prime|$. We can replace $p$ by $p^\prime + c$ to generate $Aut(G)$ such that $K(Aut(G))=p^\prime+c$, where $c$ is a constant independent of $G^\prime$ that represents the size of the shortest program that generates $Auth(G)$, given any $G$. Then we have it that $|K(Aut(G))-K(G)| < c$, which is contrary to the assumption.
\end{proof}
\end{theor}

The number of Borel-normal numbers that can be used as the degree sequence of a graph is determined by the necessary and sufficient conditions in~\cite{kim,kim2} and is numerable infinite.

\subsection{Degree-sequence targeted Entropy-deceiving graph construction}

Taking advantage of the correlation between 2 variables $X_1$, $X_2$ (starting independently) with the same probability distribution, let $M$ be a $2\times 2$ matrix with rows normalized to 1. Consider the random variables $Y_1$, $Y_2$ which satisfy

$$(Y1, Y2) = (M.X1, M.X2)$$

The correlation between $Y_1$ and $Y_2$ is just the inner product between the two rows of $M$. This can be used to generate a degree distribution of a graph with any particular Entropy, provided the resulting degree sequence complies with or is completed according to the necessary and sufficient conditions for building a graph~\cite{kim,kim2}.

\section{Graph Entropy versus Graph Algorithmic Complexity}

The ensemble of the graphs compatible with the ZK graph for the Entropy of its degree distribution consists thus of the set of networks that have near-maximal degree sequence, as the sequence distribution is uninformative (nearly every degree appears only once) and thus does not reduce statistical uncertainty, despite the algorithmic nature of the ZK graph (and assuming one does not know that the graph is deterministically generated, a reasonable assumption of ignorance characteristic of the general observer in a typical, realistic case). The size of the ensemble is thereby close to $|d|!$, the number of permutations of the elements of the degree distribution $d$ of the ZK graph, constrained by the number of sequences that can actually construct a graph~\cite{kim,kim2}. This means that, without loss of generality, any Entropy-based measure (in this case applied to the degree sequence) will be misleading, assigning high randomness after a large ensemble of equally high Entropy values when it is in fact a simple recursive graph, and thereby illustrating the limits of classical information theory for graph profiling.

\subsection{Algorithmic complexity invariance vis-\'a-vis full object description}

While this paper does not focus on alternatives to graph Entropy, alternative and complementary directions for exploring robust (if semi-computable) approaches to graph complexity have been introduced~\cite{zenilgraph}, together with numerical methods showing that one can not only robustly define the algorithmic complexity (even when semi-computable) of labelled graphs more independently of description language, but also of unlabelled graphs, as set forth in~\cite{zenilmethodsbiology}, in particular:

\begin{definition}{Algorithmic Complexity of unlabelled graphs:}
Let $D(G)$ be a lossless description of $G$ and $Aut(G)$ its automorphism group. Then, 
$$C(G)=\min\{C(D(G)) | D(G)\in S(Aut(G))\}$$
\end{definition}

\noindent where $C(G)$ is the algorithmic (Kolmogorov-Chaitin) complexity of the graph $G$ as introduced in~\cite{zenilgraph,zenilmethodsbiology} (the shortest computer program that produces $G$ upon halting) and $S(Aut(G))$ is the set of all $D$ descriptions for all graphs in $Aut(G)$, independent of $D$ (per the Invariance theorem). Which, unlike graph Entropy, is robust~\cite{zenilmethodsbiology}. In~\cite{zenilgraph}, it was in fact shown that the algorithmic complexity estimation of a labelled graph is a good approximation of the algorithmic complexity of the graph automorphism group (i.e. the unlabelled graph complexity), and is correlated in one direction to the automorphism group count.

\subsection{The fragility of Entropy and computable measures vis-\'a-vis object description}

In contrast to algorithmic complexity, no computable measure of complexity can test for all (Turing) computable regularities in a dataset~\cite{martinlof}. That is, there is no test that can be implemented as a Turing machine that takes the data as input and indicates whether it has a regularity upon halting (regularities such as ``every 5th place is occupied by a consecutive prime number'', to mention one example among an infinite number of possibilities). 

\begin{definition}
A computable regularity is a regularity for which a test can be set as a computer program running on a specific-purpose Turing machine testing for the said regularity.
\end{definition}

Common statistical tests, for example, are computable because they are designed to be effective, but no computable \emph{universal} measure of complexity can test for every computable regularity. In other words, for every computable measure capturing a data feature $X$ intended to quantify the random content of the data, one can devise a mechanistic procedure producing $X$ that deceptively simulates the said measure for all other features. 

Moreover, for every effective feature, one can devise/conceive an effective measure to test for it, but there is no computable measure able to implement a universal statistical test~\cite{martinlof}. This means that for every effective (computable) property/feature $X$ of a computable object $S$, there is a computable measure $T$ to test for $X$ in $S$ (or any object like $S$), but no computable measure $T^\prime$ exists to test for every feature $X$ in $S$ (and all the effectively enumerable computable objects like $S$).

Let $D(G)$ be a lossless description of an object $G$, meaning that $G$ can be reconstructed from $D(G)$ without any loss of information. Then there is no essential distinction between $D$ and $G$ from the algorithmic point of view because $C(G)=C(D(G)) + c $, where $c$ is the length of the translation program (in bits) between $D$ and $G$.\\

\begin{theor}
For a computable measure $H$, such as Shannon Entropy, there is no constant $c$ or logarithmic term such that $\forall G$, $|H(G_{D_1}) - H(G_{D_2})| < c$ or $|H(G_{D_1}) - H(G_{D_2})| < \log |G|$ bounding the difference as a function of the size of $G$.
\end{theor}

In other words, as we have proven by exhibiting a counter-example (the ZK graph), the Shannon Entropy $H$ of an object may diverge when applied to different lossless descriptions of the same object and cannot therefore be considered a robust measure of complexity. A measure of complexity should thus look not for a single property $X$ in (\textit{any} possible) object $S$ but for potentially an unbounded (and potentially unidentified)  number of possible properties $X= x_0, x_1, \ldots$ in any object $S$. 

A sound characterization of a complexity measure can thus be established as a function that captures strictly more information about (any) $S$ than any (computable) function. All computable functions are thus not good candidates for universal measures of complexity as they can be replaced by a measure as a function of the property (or combination) of properties of interest and nothing else.

\subsection{Dependence on assumed distributions}

An argument against the claim that Entropy yields contradictory values when used to profile randomness (even statistical randomness) is that one can change the domain of the Entropy measure in such a way as to make Entropy consistent with any possible description of a graph. For example, because we have proven that the ZK algorithm is deterministic and can only produce a single ZK graph, it follows that there is no uncertainty in the production of the object, there being only one graph for the formula. In this way, building a distribution of all formulae generating the ZK graph will always lead to Shannon Entropy $H(ZK)=0$ for the `right' description using the `right' ensemble containing only the ZK formula(e).

According to the same argument the digits of the mathematical constant $\pi$ (to mention only the most trivial example) would have Shannon Entropy $H(\pi)=0$, because the digits are produced deterministically and the `right' ensemble for $\pi$ should be that containing only formulae deterministically generating the digits $\pi$. 

Directly changing the ensemble on which Entropy operates for a specific object only facilitates conformity to some arbitrary Entropy value dictated by an arbitrary expectation, e.g. that $H(\pi_n)=0$ for any initial segment of $\pi$ of length $n$ (entailing an Entropy rate of 0 as well) because $\pi$ is deterministic and therefore no digit is surprising at all, or alternatively, $\lim_{n\rightarrow \infty} H(\pi_n) = \infty$ if Shannon Entropy is supposed to measure statistical randomness. Moreover, this misbehaviour has to do not with a lack of knowledge but with the lack of an \textit{invariance theorem}, because $\pi$ is deterministically generated and hence its digits do not fundamentally reduce uncertainty. But if one assumes that the digits of $\pi$ are not stochastic in order to assign it a Shannon Entropy equal to zero, then one is forced to concede that even perfect statistical randomness, produced by a supposedly Borel-normal number, has, in objective terms, a Shannon Entropy (and Entropy rate) equal to zero, but the highest Shannon Entropy (and Entropy rate) from an observer perspective (as it will never be certain that the streaming digits are truly $\pi$). In other words, the asymptotic behaviour after taking into consideration the digits of $\pi$ approximates maximum Shannon Entropy, but $\pi$ itself has a Shannon Entropy of zero. 

\subsection{An algorithmic Maximum Entropy Model}

Following the statistical mechanics approach~\cite{bianconi2007entropy}, a typical recursively generated graph such as the ZK graph would, based on its degree sequence, be characterized as being typically random from the observer perspective--because Shannon Entropy will find the graph to be statistically random and thus just as random as any member of the set of all  graphs with (near) maximal degree sequence Entropy--thus giving no indication of the actual recursive nature of the ZK graph and misleading the observer.

 In contrast, the type of approach introduced in~\cite{zenilgraph}, based upon trying to find clues to the recursive nature of an object such as a graph, would asymptotically find the causal nature of a recursively-generating object such as the ZK graph, independent of probability distributions, even if it is more difficult to estimate.

Rectifying the approaches based on models of maximum entropy involves updating and replacing the assumption of the maximum entropy ensemble. An example illustrating how to achieve this in the context of, e.g., a Bayesian approach, has been provided in~\cite{algorithmiccognition1} and consists in replacing the uninformative prior by the uninformative algorithmic probability distribution, the so-called Universal Distribution, as introduced by Levin~\cite{levin}. The general approach has already delivered some important results~\cite{algorithmiccognition2} by, e.g., quantifying the degree of human cognitive randomness that previous statistical approaches and measures such as Entropy made it impossible to quantify. Animated videos have been made available explaining applications to graph complexity (\url{https://youtu.be/E238zKsPCgk}) and to cognition in the context of random generation tasks (\url{https://youtu.be/E-YjBE5qm7c}). A tool has also been placed online (\url{http://complexitycalculator.com/}) for sequences and arrays, and thus the reader can experiment with an actual numerical tool and explore the differences between the statistical and the algorithmic approaches.

\section{Conclusions}
\label{conclusion}

The methods introduced here allow the construction of `Borel-normal pseudo-random graphs', uncomputable number-based graphs and algorithmically produced graphs, while illustrating the shortcomings of computable graph-theoretic and Entropy approaches to graph complexity beyond random feature selection, and their failure when it comes to profiling  randomness and hence causal-content (as opposed to randomness). 

We have shown that Entropy is highly observer-dependent even in the face of full accuracy and access to lossless object descriptions and thus has to be complemented by measures of algorithmic content. We have produced specific complexity-deceiving graphs for which Entropy retrieves disparate values when an object is described differently (thus with different underlying distributions), even when the descriptions reconstruct exactly the same, and only the same, object. This drawback of Shannon Entropy, ultimately related to its dependence on distribution, is all the more serious because it is easily overlooked in the case of objects other than strings, for instance, graphs. For an object such as a graph, we have shown that changing the descriptions may not only change the values but actually produce divergent, contradictory values. 

We constructed a graph ZK about which the following is true when it is described by its adjacency matrix $Adj(ZK)$: $\lim_{n\rightarrow \infty}H(Adj(ZK_n))=0$ for growing graph size $n$. Contradictorily, considering the same ZK graph degree sequence, we found that $\lim_{n\rightarrow \infty}H(Seq(ZK_n))=\infty$ for the same growth rate $n$, even though both $Adj(G_n)$ and $Seq(G_n)$ are lossless descriptions of the same graph that construct exactly the same ZK graph, and only a ZK graph.

This means that not only does one need to choose a description of interest in order to apply a definition of Entropy, such as the adjacency matrix of a network (or its incidence or Laplacian) or its degree sequence, but that as soon as the choice is made, Entropy becomes a trivial counting function of the specific feature of interest, and of that feature alone. In the case of, for example, the adjacency matrix of a network (or any related matrix associated with the graph, such as the incidence or Laplacian matrices), Entropy becomes a function of edge density, while for degree sequence, Entropy becomes a function of sequence normality. Entropy can thus trivially be replaced by such functions without any loss, but it cannot be used to profile the object (randomness, or information content) in any way independent of an arbitrary feature of interest. 

These results and observations have far-reaching consequences. For example, recent literature appears contradictory, by turns suggesting that cancer cells display an increase in Entropy~\cite{teschendorff}, and also reporting that cancer cells display a decrease in Entropy~\cite{west}, in both cases applied to a function of degree distribution over networks of molecular interactions. Cells are also believed to be in a state of criticality between evolvability and robustness~\cite{aldana,csermely} that may make them look random though they are not. This means that Entropy may be overestimating randomness in the best case or misleading in the worst case, as we have found in the instance of disparate values for the same objects, thus suggesting that additional safeguards are needed to achieve consistency and soundness. 

New developments~\cite{zenilgraph,zenilmethodsbiology} promise more robust complementary measures of (graph) complexity less dependent on object description, measures based upon the mathematical theory of randomness and algorithmic probability which are better equipped to profile causality and algorithmic information content and cover statistical randomness and thus can be considered an observer-improved generalization of Shannon Entropy.


\begin{thebibliography}{23}%
\makeatletter
\providecommand \@ifxundefined [1]{%
 \@ifx{#1\undefined}
}%
\providecommand \@ifnum [1]{%
 \ifnum #1\expandafter \@firstoftwo
 \else \expandafter \@secondoftwo
 \fi
}%
\providecommand \@ifx [1]{%
 \ifx #1\expandafter \@firstoftwo
 \else \expandafter \@secondoftwo
 \fi
}%
\providecommand \natexlab [1]{#1}%
\providecommand \enquote  [1]{``#1''}%
\providecommand \bibnamefont  [1]{#1}%
\providecommand \bibfnamefont [1]{#1}%
\providecommand \citenamefont [1]{#1}%
\providecommand \href@noop [0]{\@secondoftwo}%
\providecommand \href [0]{\begingroup \@sanitize@url \@href}%
\providecommand \@href[1]{\@@startlink{#1}\@@href}%
\providecommand \@@href[1]{\endgroup#1\@@endlink}%
\providecommand \@sanitize@url [0]{\catcode `\\12\catcode `\$12\catcode
  `\&12\catcode `\#12\catcode `\^12\catcode `\_12\catcode `\%12\relax}%
\providecommand \@@startlink[1]{}%
\providecommand \@@endlink[0]{}%
\providecommand \url  [0]{\begingroup\@sanitize@url \@url }%
\providecommand \@url [1]{\endgroup\@href {#1}{\urlprefix }}%
\providecommand \urlprefix  [0]{URL }%
\providecommand \Eprint [0]{\href }%
\providecommand \doibase [0]{http://dx.doi.org/}%
\providecommand \selectlanguage [0]{\@gobble}%
\providecommand \bibinfo  [0]{\@secondoftwo}%
\providecommand \bibfield  [0]{\@secondoftwo}%
\providecommand \translation [1]{[#1]}%
\providecommand \BibitemOpen [0]{}%
\providecommand \bibitemStop [0]{}%
\providecommand \bibitemNoStop [0]{.\EOS\space}%
\providecommand \EOS [0]{\spacefactor3000\relax}%
\providecommand \BibitemShut  [1]{\csname bibitem#1\endcsname}%
\let\auto@bib@innerbib\@empty
\bibitem [{\citenamefont {Milo}\ \emph {et~al.}(2002)\citenamefont {Milo},
  \citenamefont {Shen-Orr}, \citenamefont {Itzkovitz}, \citenamefont {Kashtan},
  \citenamefont {Chklovskii},\ and\ \citenamefont {Alon}}]{alon1}%
  \BibitemOpen
  \bibfield  {author} {\bibinfo {author} {\bibfnamefont {R.}~\bibnamefont
  {Milo}}, \bibinfo {author} {\bibfnamefont {S.}~\bibnamefont {Shen-Orr}},
  \bibinfo {author} {\bibfnamefont {S.}~\bibnamefont {Itzkovitz}}, \bibinfo
  {author} {\bibfnamefont {N.}~\bibnamefont {Kashtan}}, \bibinfo {author}
  {\bibfnamefont {D.}~\bibnamefont {Chklovskii}}, \ and\ \bibinfo {author}
  {\bibfnamefont {U.}~\bibnamefont {Alon}},\ }\href@noop {} {\bibfield
  {journal} {\bibinfo  {journal} {Science}\ }\textbf {\bibinfo {volume}
  {298}},\ \bibinfo {pages} {824} (\bibinfo {year} {2002})}\BibitemShut
  {NoStop}%
\bibitem [{\citenamefont {Shen-Orr}\ \emph {et~al.}(2002)\citenamefont
  {Shen-Orr}, \citenamefont {Milo}, \citenamefont {Mangan},\ and\ \citenamefont
  {Alon}}]{alon2}%
  \BibitemOpen
  \bibfield  {author} {\bibinfo {author} {\bibfnamefont {S.~S.}\ \bibnamefont
  {Shen-Orr}}, \bibinfo {author} {\bibfnamefont {R.}~\bibnamefont {Milo}},
  \bibinfo {author} {\bibfnamefont {S.}~\bibnamefont {Mangan}}, \ and\ \bibinfo
  {author} {\bibfnamefont {U.}~\bibnamefont {Alon}},\ }\href@noop {} {\bibfield
   {journal} {\bibinfo  {journal} {Nature genetics}\ }\textbf {\bibinfo
  {volume} {31}},\ \bibinfo {pages} {64} (\bibinfo {year} {2002})}\BibitemShut
  {NoStop}%
\bibitem [{\citenamefont {Lu}\ \emph {et~al.}(2015)\citenamefont {Lu},
  \citenamefont {Li},\ and\ \citenamefont {Wang}}]{lu}%
  \BibitemOpen
  \bibfield  {author} {\bibinfo {author} {\bibfnamefont {G.}~\bibnamefont
  {Lu}}, \bibinfo {author} {\bibfnamefont {B.}~\bibnamefont {Li}}, \ and\
  \bibinfo {author} {\bibfnamefont {L.}~\bibnamefont {Wang}},\ }\href@noop {}
  {\bibfield  {journal} {\bibinfo  {journal} {Entropy}\ }\textbf {\bibinfo
  {volume} {17}},\ \bibinfo {pages} {8217} (\bibinfo {year}
  {2015})}\BibitemShut {NoStop}%
\bibitem [{\citenamefont {Shannon}(1948)}]{shannon}%
  \BibitemOpen
  \bibfield  {author} {\bibinfo {author} {\bibfnamefont {C.}~\bibnamefont
  {Shannon}},\ }\href@noop {} {\bibfield  {journal} {\bibinfo  {journal}
  {Mathematical Reviews (MathSciNet): MR10, 133e}\ } (\bibinfo {year}
  {1948})}\BibitemShut {NoStop}%
\bibitem [{\citenamefont {Solomonoff}(1964)}]{solomonoff}%
  \BibitemOpen
  \bibfield  {author} {\bibinfo {author} {\bibfnamefont {R.~J.}\ \bibnamefont
  {Solomonoff}},\ }\href@noop {} {\bibfield  {journal} {\bibinfo  {journal}
  {Information and control}\ }\textbf {\bibinfo {volume} {7}},\ \bibinfo
  {pages} {1} (\bibinfo {year} {1964})}\BibitemShut {NoStop}%
\bibitem [{\citenamefont {Kolmogorov}(1968)}]{kolmogorov}%
  \BibitemOpen
  \bibfield  {author} {\bibinfo {author} {\bibfnamefont {A.~N.}\ \bibnamefont
  {Kolmogorov}},\ }\href@noop {} {\bibfield  {journal} {\bibinfo  {journal}
  {International Journal of Computer Mathematics}\ }\textbf {\bibinfo {volume}
  {2}},\ \bibinfo {pages} {157} (\bibinfo {year} {1968})}\BibitemShut {NoStop}%
\bibitem [{\citenamefont {Chaitin}(1966)}]{chaitin}%
  \BibitemOpen
  \bibfield  {author} {\bibinfo {author} {\bibfnamefont {G.~J.}\ \bibnamefont
  {Chaitin}},\ }\href@noop {} {\bibfield  {journal} {\bibinfo  {journal}
  {Journal of the ACM (JACM)}\ }\textbf {\bibinfo {volume} {13}},\ \bibinfo
  {pages} {547} (\bibinfo {year} {1966})}\BibitemShut {NoStop}%
\bibitem [{\citenamefont {Zenil}\ \emph {et~al.}(2014)\citenamefont {Zenil},
  \citenamefont {Soler-Toscano}, \citenamefont {Dingle},\ and\ \citenamefont
  {Louis}}]{zenilgraph}%
  \BibitemOpen
  \bibfield  {author} {\bibinfo {author} {\bibfnamefont {H.}~\bibnamefont
  {Zenil}}, \bibinfo {author} {\bibfnamefont {F.}~\bibnamefont
  {Soler-Toscano}}, \bibinfo {author} {\bibfnamefont {K.}~\bibnamefont
  {Dingle}}, \ and\ \bibinfo {author} {\bibfnamefont {A.~A.}\ \bibnamefont
  {Louis}},\ }\href@noop {} {\bibfield  {journal} {\bibinfo  {journal} {Physica
  A: Statistical Mechanics and its Applications}\ }\textbf {\bibinfo {volume}
  {404}},\ \bibinfo {pages} {341} (\bibinfo {year} {2014})}\BibitemShut
  {NoStop}%
\bibitem [{\citenamefont {Zenil}\ \emph
  {et~al.}(2016{\natexlab{a}})\citenamefont {Zenil}, \citenamefont {Kiani},\
  and\ \citenamefont {Tegn{\'e}r}}]{zenilmethodsbiology}%
  \BibitemOpen
  \bibfield  {author} {\bibinfo {author} {\bibfnamefont {H.}~\bibnamefont
  {Zenil}}, \bibinfo {author} {\bibfnamefont {N.~A.}\ \bibnamefont {Kiani}}, \
  and\ \bibinfo {author} {\bibfnamefont {J.}~\bibnamefont {Tegn{\'e}r}},\ }in\
  \href@noop {} {\emph {\bibinfo {booktitle} {Seminars in cell \& developmental
  biology}}},\ Vol.~\bibinfo {volume} {51}\ (\bibinfo {organization}
  {Elsevier},\ \bibinfo {year} {2016})\ pp.\ \bibinfo {pages}
  {32--43}\BibitemShut {NoStop}%
\bibitem [{\citenamefont {Zenil}\ \emph {et~al.}(2015)\citenamefont {Zenil},
  \citenamefont {Soler-Toscano}, \citenamefont {Delahaye},\ and\ \citenamefont
  {Gauvrit}}]{kolmo2d}%
  \BibitemOpen
  \bibfield  {author} {\bibinfo {author} {\bibfnamefont {H.}~\bibnamefont
  {Zenil}}, \bibinfo {author} {\bibfnamefont {F.}~\bibnamefont
  {Soler-Toscano}}, \bibinfo {author} {\bibfnamefont {J.-P.}\ \bibnamefont
  {Delahaye}}, \ and\ \bibinfo {author} {\bibfnamefont {N.}~\bibnamefont
  {Gauvrit}},\ }\href@noop {} {\bibfield  {journal} {\bibinfo  {journal} {PeerJ
  Computer Science}\ }\textbf {\bibinfo {volume} {1}},\ \bibinfo {pages} {e23}
  (\bibinfo {year} {2015})}\BibitemShut {NoStop}%
\bibitem [{\citenamefont {Mowshowitz}\ and\ \citenamefont
  {Dehmer}(2012)}]{mowshowitz}%
  \BibitemOpen
  \bibfield  {author} {\bibinfo {author} {\bibfnamefont {A.}~\bibnamefont
  {Mowshowitz}}\ and\ \bibinfo {author} {\bibfnamefont {M.}~\bibnamefont
  {Dehmer}},\ }\href@noop {} {\bibfield  {journal} {\bibinfo  {journal}
  {Entropy}\ }\textbf {\bibinfo {volume} {14}},\ \bibinfo {pages} {559}
  (\bibinfo {year} {2012})}\BibitemShut {NoStop}%
\bibitem [{\citenamefont {Simonyi}(1995)}]{simonyi}%
  \BibitemOpen
  \bibfield  {author} {\bibinfo {author} {\bibfnamefont {G.}~\bibnamefont
  {Simonyi}},\ }\href@noop {} {\bibfield  {journal} {\bibinfo  {journal}
  {Combinatorial Optimization}\ }\textbf {\bibinfo {volume} {20}},\ \bibinfo
  {pages} {399} (\bibinfo {year} {1995})}\BibitemShut {NoStop}%
\bibitem [{\citenamefont {Dehmer}\ and\ \citenamefont
  {Mowshowitz}(2011)}]{mowshowitz2}%
  \BibitemOpen
  \bibfield  {author} {\bibinfo {author} {\bibfnamefont {M.}~\bibnamefont
  {Dehmer}}\ and\ \bibinfo {author} {\bibfnamefont {A.}~\bibnamefont
  {Mowshowitz}},\ }\href@noop {} {\bibfield  {journal} {\bibinfo  {journal}
  {Information Sciences}\ }\textbf {\bibinfo {volume} {181}},\ \bibinfo {pages}
  {57} (\bibinfo {year} {2011})}\BibitemShut {NoStop}%
\bibitem [{\citenamefont {Zenil}(2013)}]{zenildata}%
  \BibitemOpen
  \bibfield  {author} {\bibinfo {author} {\bibfnamefont {H.}~\bibnamefont
  {Zenil}},\ }\href@noop {} {\bibfield  {journal} {\bibinfo  {journal}
  {Predictability in the world: philosophy and science in the complex world of
  Big Data, Springer Verlag (forthcoming)}\ } (\bibinfo {year}
  {2013})}\BibitemShut {NoStop}%
\bibitem [{\citenamefont {Zenil}\ \emph
  {et~al.}(2016{\natexlab{b}})\citenamefont {Zenil}, \citenamefont
  {Soler-Toscano}, \citenamefont {Kiani}, \citenamefont {Hern\'andez-Orozco},\
  and\ \citenamefont {Rueda-Toicen}}]{zenilbdm}%
  \BibitemOpen
  \bibfield  {author} {\bibinfo {author} {\bibfnamefont {H.}~\bibnamefont
  {Zenil}}, \bibinfo {author} {\bibfnamefont {F.}~\bibnamefont
  {Soler-Toscano}}, \bibinfo {author} {\bibfnamefont {N.~A.}\ \bibnamefont
  {Kiani}}, \bibinfo {author} {\bibfnamefont {S.}~\bibnamefont
  {Hern\'andez-Orozco}}, \ and\ \bibinfo {author} {\bibfnamefont
  {A.}~\bibnamefont {Rueda-Toicen}},\ }\href@noop {} {\bibfield  {journal}
  {\bibinfo  {journal} {submitted}\ } (\bibinfo {year}
  {2016}{\natexlab{b}})}\BibitemShut {NoStop}%
\bibitem [{\citenamefont {Levin}(1974)}]{levin}%
  \BibitemOpen
  \bibfield  {author} {\bibinfo {author} {\bibfnamefont {L.~A.}\ \bibnamefont
  {Levin}},\ }\href@noop {} {\bibfield  {journal} {\bibinfo  {journal}
  {Problemy Peredachi Informatsii}\ }\textbf {\bibinfo {volume} {10}},\
  \bibinfo {pages} {30} (\bibinfo {year} {1974})}\BibitemShut {NoStop}%
\bibitem [{\citenamefont {Calude}\ \emph {et~al.}(2002)\citenamefont {Calude},
  \citenamefont {Dinneen}, \citenamefont {Shu} \emph {et~al.}}]{calude}%
  \BibitemOpen
  \bibfield  {author} {\bibinfo {author} {\bibfnamefont {C.~S.}\ \bibnamefont
  {Calude}}, \bibinfo {author} {\bibfnamefont {M.~J.}\ \bibnamefont {Dinneen}},
  \bibinfo {author} {\bibfnamefont {C.-K.}\ \bibnamefont {Shu}},  \emph
  {et~al.},\ }\href@noop {} {\bibfield  {journal} {\bibinfo  {journal}
  {Experimental Mathematics}\ }\textbf {\bibinfo {volume} {11}},\ \bibinfo
  {pages} {361} (\bibinfo {year} {2002})}\BibitemShut {NoStop}%
\bibitem [{\citenamefont {Martin-L{\"o}f}(1966)}]{martinlof}%
  \BibitemOpen
  \bibfield  {author} {\bibinfo {author} {\bibfnamefont {P.}~\bibnamefont
  {Martin-L{\"o}f}},\ }\href@noop {} {\bibfield  {journal} {\bibinfo  {journal}
  {Information and control}\ }\textbf {\bibinfo {volume} {9}},\ \bibinfo
  {pages} {602} (\bibinfo {year} {1966})}\BibitemShut {NoStop}%
\bibitem [{\citenamefont {Barab{\'a}si}\ and\ \citenamefont
  {Albert}(1999)}]{barabasi}%
  \BibitemOpen
  \bibfield  {author} {\bibinfo {author} {\bibfnamefont {A.-L.}\ \bibnamefont
  {Barab{\'a}si}}\ and\ \bibinfo {author} {\bibfnamefont {R.}~\bibnamefont
  {Albert}},\ }\href@noop {} {\bibfield  {journal} {\bibinfo  {journal}
  {science}\ }\textbf {\bibinfo {volume} {286}},\ \bibinfo {pages} {509}
  (\bibinfo {year} {1999})}\BibitemShut {NoStop}%
\bibitem [{\citenamefont {Sol{\'e}}\ and\ \citenamefont
  {Valverde}(2008)}]{sole}%
  \BibitemOpen
  \bibfield  {author} {\bibinfo {author} {\bibfnamefont {R.~V.}\ \bibnamefont
  {Sol{\'e}}}\ and\ \bibinfo {author} {\bibfnamefont {S.}~\bibnamefont
  {Valverde}},\ }\href@noop {} {\bibfield  {journal} {\bibinfo  {journal}
  {Journal of The Royal Society Interface}\ }\textbf {\bibinfo {volume} {5}},\
  \bibinfo {pages} {129} (\bibinfo {year} {2008})}\BibitemShut {NoStop}%
\bibitem [{\citenamefont {Champernowne}(1933)}]{champernowne}%
  \BibitemOpen
  \bibfield  {author} {\bibinfo {author} {\bibfnamefont {D.~G.}\ \bibnamefont
  {Champernowne}},\ }\href@noop {} {\bibfield  {journal} {\bibinfo  {journal}
  {Journal of the London Mathematical Society}\ }\textbf {\bibinfo {volume}
  {1}},\ \bibinfo {pages} {254} (\bibinfo {year} {1933})}\BibitemShut {NoStop}%
\bibitem [{\citenamefont {{\'E}mile~Borel}(1909)}]{borel}%
  \BibitemOpen
  \bibfield  {author} {\bibinfo {author} {\bibfnamefont {M.}~\bibnamefont
  {{\'E}mile~Borel}},\ }\href@noop {} {\bibfield  {journal} {\bibinfo
  {journal} {Rendiconti del Circolo Matematico di Palermo (1884-1940)}\
  }\textbf {\bibinfo {volume} {27}},\ \bibinfo {pages} {247} (\bibinfo {year}
  {1909})}\BibitemShut {NoStop}%
\bibitem [{\citenamefont {Kim}\ \emph {et~al.}(2009)\citenamefont {Kim},
  \citenamefont {Toroczkai}, \citenamefont {Erd{\H{o}}s}, \citenamefont
  {Mikl{\'o}s},\ and\ \citenamefont {Sz{\'e}kely}}]{kim}%
  \BibitemOpen
  \bibfield  {author} {\bibinfo {author} {\bibfnamefont {H.}~\bibnamefont
  {Kim}}, \bibinfo {author} {\bibfnamefont {Z.}~\bibnamefont {Toroczkai}},
  \bibinfo {author} {\bibfnamefont {P.~L.}\ \bibnamefont {Erd{\H{o}}s}},
  \bibinfo {author} {\bibfnamefont {I.}~\bibnamefont {Mikl{\'o}s}}, \ and\
  \bibinfo {author} {\bibfnamefont {L.~A.}\ \bibnamefont {Sz{\'e}kely}},\
  }\href@noop {} {\bibfield  {journal} {\bibinfo  {journal} {Journal of Physics
  A: Mathematical and Theoretical}\ }\textbf {\bibinfo {volume} {42}},\
  \bibinfo {pages} {392001} (\bibinfo {year} {2009})}\BibitemShut {NoStop}%
\end{thebibliography}%


\begin{thebibliography}{10}

\bibitem{barabasi}
Albert-L{\'a}szl{\'o} Barab{\'a}si and R{\'e}ka Albert.
\newblock Emergence of scaling in random networks.
\newblock {\em science}, 286(5439):509--512, 1999.

\bibitem{bianconi2007entropy}
Ginestra Bianconi.
\newblock The entropy of randomized network ensembles.
\newblock {\em EPL (Europhysics Letters)}, 81(2):28005, 2007.

\bibitem{boccaletti}
S.~Boccaletti et~al.
\newblock The structure and dynamics of multilayer networks.
\newblock {\em Physics Reports}, 544(1):1--122, 2014.

\bibitem{borel}
E~Borel.
\newblock Les probabilit{\'e}s d{\'e}nombrables et leurs applications
  arithm{\'e}tiques.
\newblock {\em Rendiconti del Circolo Matematico di Palermo (1884-1940)},
  27(1):247--271, 1909.

\bibitem{calude}
Cristian~S Calude, Michael~J Dinneen, Chi-Kou Shu, et~al.
\newblock Computing a glimpse of randomness.
\newblock {\em Experimental Mathematics}, 11(3):361--370, 2002.

\bibitem{chaitin}
Gregory~J Chaitin.
\newblock On the length of programs for computing finite binary sequences.
\newblock {\em Journal of the ACM (JACM)}, 13(4):547--569, 1966.

\bibitem{champernowne}
David~G Champernowne.
\newblock The construction of decimals normal in the scale of ten.
\newblock {\em Journal of the London Mathematical Society}, 1(4):254--260,
  1933.

\bibitem{chen2014entropy}
Zengqiang Chen, Matthias Dehmer, Frank Emmert-Streib, and Yongtang Shi.
\newblock Entropy bounds for dendrimers.
\newblock {\em Applied Mathematics and Computation}, 242:462--472, 2014.

\bibitem{csermely}
Peter Csermely et~al.
\newblock Cancer stem cells display extremely large evolvability: alternating
  plastic and rigid networks as a potential mechanism.
\newblock {\em Seminars in Cancer Biology}, 30:42--51, 2015.

\bibitem{dehmer2008entropy}
Matthias Dehmer, Stephan Borgert, and Frank Emmert-Streib.
\newblock Entropy bounds for hierarchical molecular networks.
\newblock {\em PLoS One}, 3(8):e3079, 2008.

\bibitem{Dehmer3}
Matthias Dehmer and Abbe Mowshowitz.
\newblock A history of graph entropy measures.
\newblock {\em Information Sciences}, 181(1):57--78, 2011.

\bibitem{erdos1960evolution}
Paul Erdos and Alfr{\'e}d R{\'e}nyi.
\newblock On the evolution of random graphs.
\newblock {\em Publ. Math. Inst. Hung. Acad. Sci}, 5(1):17--60, 1960.

\bibitem{Estrada2}
Ernesto Estrada.
\newblock Quantifying network heterogeneity.
\newblock {\em Physical Review E}, 82(6):066102, 2010.

\bibitem{estrada2014walk}
Ernesto Estrada, A~Jos{\'e}, and Naomichi Hatano.
\newblock Walk entropies in graphs.
\newblock {\em Linear Algebra and its Applications}, 443:235--244, 2014.

\bibitem{orsini}
Orsini et~al.
\newblock Quantifying randomness in real networks.
\newblock {\em Nature Communications}, 6:8627, 2015.

\bibitem{kim2}
H.~Kim, C.~I. Del~Genio, K.~E. Bassler, and Z.~Toroczkai.
\newblock Degree-based graph construction.
\newblock {\em New Journal of Physics}, 14:023012, 2012.

\bibitem{kim}
Hyunju Kim, Zolt{\'a}n Toroczkai, P{\'e}ter~L Erd{\H{o}}s, Istv{\'a}n
  Mikl{\'o}s, and L{\'a}szl{\'o}~A Sz{\'e}kely.
\newblock Degree-based graph construction.
\newblock {\em Journal of Physics A: Mathematical and Theoretical},
  42(39):392001, 2009.

\bibitem{kolmogorov}
Andrei~Nikolaevich Kolmogorov.
\newblock Three approaches to the quantitative definition of information*.
\newblock {\em International Journal of Computer Mathematics}, 2(1-4):157--168,
  1968.

\bibitem{korner1988random}
Janos Korner and Katalin Marton.
\newblock Random access communication and graph entropy.
\newblock {\em IEEE transactions on information theory}, 34(2):312--314, 1988.

\bibitem{levin}
Leonid~A Levin.
\newblock Laws of information conservation (nongrowth) and aspects of the
  foundation of probability theory.
\newblock {\em Problemy Peredachi Informatsii}, 10(3):30--35, 1974.

\bibitem{li}
Ming Li and Paul Vit{\'a}nyi.
\newblock {\em An introduction to Kolmogorov complexity and its applications}.
\newblock Springer Science \& Business Media, 2009.

\bibitem{lu}
Guoxiang Lu, Bingqing Li, and Lijia Wang.
\newblock Some new properties for degree-based graph entropies.
\newblock {\em Entropy}, 17(12):8217--8227, 2015.

\bibitem{martinlof}
Per Martin-L{\"o}f.
\newblock The definition of random sequences.
\newblock {\em Information and control}, 9(6):602--619, 1966.

\bibitem{alon1}
Ron Milo, Shai Shen-Orr, Shalev Itzkovitz, Nadav Kashtan, Dmitri Chklovskii,
  and Uri Alon.
\newblock Network motifs: simple building blocks of complex networks.
\newblock {\em Science}, 298(5594):824--827, 2002.

\bibitem{mowshowitz1968entropy}
Abbe Mowshowitz.
\newblock Entropy and the complexity of graphs: I. an index of the relative
  complexity of a graph.
\newblock {\em The bulletin of mathematical biophysics}, 30(1):175--204, 1968.

\bibitem{mowshowitz}
Abbe Mowshowitz and Matthias Dehmer.
\newblock Entropy and the complexity of graphs revisited.
\newblock {\em Entropy}, 14(3):559--570, 2012.

\bibitem{sengupta2016application}
Dipendra~C Sengupta and Jharna~D Sengupta.
\newblock Application of graph entropy in CRISPR and repeats detection in DNA
  sequences.
\newblock {\em Computational Molecular Bioscience}, 6(03):41, 2016.

\bibitem{shannon}
CE~Shannon.
\newblock A mathematical theory of communication, {B}ell {S}ystem technical
  journal 27: 379-423 and 623--656.
\newblock {\em Mathematical Reviews (MathSciNet): MR10, 133e}, 1948.

\bibitem{alon2}
Shai~S Shen-Orr, Ron Milo, Shmoolik Mangan, and Uri Alon.
\newblock Network motifs in the transcriptional regulation network of
  Escherichia coli.
\newblock {\em Nature genetics}, 31(1):64--68, 2002.

\bibitem{simonyi}
G{\'a}bor Simonyi.
\newblock Graph entropy: a survey.
\newblock {\em Combinatorial Optimization}, 20:399--441, 1995.

\bibitem{sole}
Ricard~V Sol{\'e} and Sergi Valverde.
\newblock Spontaneous emergence of modularity in cellular networks.
\newblock {\em Journal of The Royal Society Interface}, 5(18):129--133, 2008.

\bibitem{ctm}
Fernando Soler-Toscano, Hector Zenil, Jean-Paul Delahaye, and Nicolas Gauvrit.
\newblock Calculating Kolmogorov complexity from the output frequency
  distributions of small Turing machines.
\newblock {\em PloS one}, 9(5):e96223, 2014.

\bibitem{solomonoff}
Ray~J Solomonoff.
\newblock A formal theory of inductive inference. part i.
\newblock {\em Information and control}, 7(1):1--22, 1964.

\bibitem{teschendorff}
Andrew~E Teschendorff and Simone Severini.
\newblock Increased entropy of signal transduction in the cancer metastasis
  phenotype.
\newblock {\em BMC Systems Biology}, 4(104), 2010.

\bibitem{aldana}
C.~Torres-Sosa, S.~Huang, and M.~Aldana.
\newblock Criticality is an emergent property of genetic networks that exhibit
  evolvability.
\newblock {\em PLoS Comput Biol}, 8(9):e1002669, 2012.

\bibitem{trucco1956note}
Ernesto Trucco.
\newblock A note on the information content of graphs.
\newblock {\em Bulletin of Mathematical Biology}, 18(2):129--135, 1956.

\bibitem{west}
James West, Simone~Severini, Ginestra~Bianconi, and Andrew~E. Teschendorff.
\newblock Differential network entropy reveals cancer system hallmarks.
\newblock {\em Scientific Reports}, 2(802), 2012.

\bibitem{zenildata}
Hector Zenil.
\newblock Small data matters, correlation versus causation and algorithmic data
  analytics.
\newblock {\em Predictability in the world: philosophy and science in the
  complex world of Big Data, Springer Verlag (forthcoming)}, 2013.

\bibitem{zenilmethodsbiology}
Hector Zenil, Narsis~A Kiani, and Jesper Tegn{\'e}r.
\newblock Methods of information theory and algorithmic complexity for network
  biology.
\newblock {\em Seminars in cell \& developmental biology}, 51:32--43, 2016.

\bibitem{kolmo2d}
Hector Zenil, Fernando Soler-Toscano, Jean-Paul Delahaye, and Nicolas Gauvrit.
\newblock Two-dimensional kolmogorov complexity and an empirical validation of
  the coding theorem method by compressibility.
\newblock {\em PeerJ Computer Science}, 1:e23, 2015.

\bibitem{zenilgraph}
Hector Zenil, Fernando Soler-Toscano, Kamaludin Dingle, and Ard~A Louis.
\newblock Correlation of automorphism group size and topological properties
  with program-size complexity evaluations of graphs and complex networks.
\newblock {\em Physica A: Statistical Mechanics and its Applications},
  404:341--358, 2014.

\bibitem{zenilbdm}
Hector Zenil, Francisco Soler-Toscano, Narsis~A. Kiani, Santiago
  Hern\'andez-Orozco, and Antonio Rueda-Toicen.
\newblock A decomposition method for global evaluation of {S}hannon entropy and
  local estimations of algorithmic complexity.
\newblock {\em 	arXiv:1609.00110 [cs.IT]}, 2016.

\bibitem{lempelziv}
Jacob Ziv and Abraham Lempel.
\newblock Compression of individual sequences via variable-rate coding.
\newblock {\em IEEE transactions on Information Theory}, 24(5):530--536, 1978.

\bibitem{mowshowitz2} M. Dehmer and A. Mowshowitz.
\newblock A history of graph entropy measures.
\newblock {\em Information Sciences} 181, 57--78, 2011.

\bibitem{algorithmiccognition1} N. Gauvrit, H. Zenil,  and J. Tegn{\'e}r.
\newblock The Information-theoretic and Algorithmic Approach toHuman, Animal and Artificial Cognition
\newblock {\em Springer Verlag} (in press)

\bibitem{algorithmiccognition2} N. Gauvrit, H. Zenil, F. Soler-Toscano, J.-P. Delahaye, and P. Bruger.
\newblock Human behavioral complexity peaks at age 25.
\newblock {\em PLoS Comput Biol.} 13:4, e1005408, 2017.


\end{thebibliography}
\end{document}